\documentclass{article}

\usepackage[preprint]{neurips_2025}  

\usepackage{graphicx, graphics}
\usepackage[table]{xcolor}
\usepackage{amsmath}
\usepackage{wrapfig}
\usepackage{booktabs}

\usepackage{colortbl}

\usepackage[utf8]{inputenc} 
\usepackage[T1]{fontenc}    
\usepackage{hyperref}       
\usepackage{url}            
\usepackage{booktabs}       
\usepackage{amsfonts}       
\usepackage{nicefrac}       
\usepackage{microtype}      
\usepackage{xcolor}         
\usepackage{soul}

\newcommand{\Qphi}{$\mathcal{Q}_\phi$}

\title{POLARIS: A High-contrast Polarimetric Imaging Benchmark Dataset for Exoplanetary Disk Representation Learning}

\author{%
Fangyi Cao\\
UC Riverside\\
\And
Bin Ren\\
OCA/MPIA\\
\And
Zihao Wang\\
UT Chattanooga/MGH/Harvard\\
\And
Shiwei Fu\\
UC Riverside\\
\And
Youbin Mo\\
UC San Diego\\
\And
Xiaoyang Liu\\
Adobe\\
\And
Yuzhou Chen\\
 UC Riverside\\
\And
Weixin Yao\\
UC Riverside
}

\begin{document}

\makeatletter
\newif\if@submission
\@submissionfalse
\makeatother

\maketitle
\newcommand*{\QED}{\hfill\ensuremath{\blacksquare}}%
\newcommand{\farcs}{\mbox{\ensuremath{.\!\!^{\prime\prime}}}}
\newcommand{\fdg}{\mbox{\ensuremath{.\!\!^\circ}}}
\newcommand{\arcsec}{^{\prime\prime}}
\newcommand{\micron}{\mu\rm{m}}
\newcommand{\aj}{The Astronomical Journal}
\newcommand{\pasp}{Publications of the Astronomical Society of the Pacific}
\newcommand{\apjl}{The Astrophysical Journal Letters}
\newcommand{\apj}{The Astrophysical Journal}
\newcommand{\apjs}{The Astrophysical Journal Supplement Series}
\newcommand{\nat}{Nature}
\newcommand{\ssr}{Space Science Reviews}
\newcommand{\procspie}{Proceedings of the SPIE}
\newcommand{\aap}{Astronomy and Astrophysics}
\newcommand{\mnras}{Monthly Notices of the Royal Astronomical Society}
\newcommand{\araa}{Annual Review of Astronomy and Astrophysics}
\newcommand{\aaps}{Astronomy and Astrophysics Supplement Series}
\newcommand{\jgr}{Journal of Geophysics Research}
\newcommand{\icarus}{Icarus}
\newcommand{\pasa}{Publications of the Astronomical Society of Australia}
\newcommand{\rmxaa}{Revista Mexicana de Astronomia y Astrofisica}
\newcommand{\jqsrt}{Journal of Quantitative Spectroscopy and Radiative Trasfer}
\newcommand{\jatis}{Journal of Astronomical Telescopes, Instruments, and Systems}

\newcommand{\Qr}{$\mathcal{Q}_\phi$}
\newcommand{\Ur}{$\mathcal{U}_\phi$}
\newcommand{\degr}{$^\circ$}
\newcommand\numberthis{\addtocounter{equation}{1}\tag{\theequation}}
\newcommand{\mdash}{---}
\newcommand{\ndash}{--}

\newcommand{\explain}{\deleted}
\newcommand{\earth}{\oplus}

\begin{abstract}

With over $10^6$ images from more than $10^4$ exposures using state-of-the-art high-contrast imagers (e.g., Gemini Planet Imager: GPI, Very Large Telescope/SPHERE) in the search for exoplanets, can the integration of artificial intelligence (AI) serve as a transformative tool in imaging Earth-like exoplanets in the upcoming decade? {In this paper, we introduce a benchmark and tackle this question from polarimetric image representation learning perspective.} In the past decade, despite extensive time and resource investment, only a handful of new exoplanets have been directly imaged. Existing exoplanet imaging approaches also heavily rely on labor-intensive labeling of reference stars, which act as background information to recover foreground circumstellar objects (either circumstellar disks or exoplanets) for target stars. 
With our POLARIS (POlarized Light dAta for total intensity Representation learning of direct Imaging of exoplanetary Systems) dataset, we classify reference star and circumstellar disk images using the entire public SPHERE/IRDIS polarized light observations collected since 2014, requiring less than 10\% manual labeling. We evaluate a range of models, including statistical models, probabilistic generative models, and state-of-the-art large vision-language models (LVLMs), and provide baseline measures for performance. We also propose an unsupervised generative representation learning framework, which integrates these models and achieves superior performance on this task, further enhancing the representational power and classification accuracy within our contrastive learning framework. To the best of our knowledge, our work introduces for the {first} time a high-quality and uniformly reduced exoplanet imaging dataset—exceedingly rare in the astrophysics community and equally scarce in machine learning domains, and we also develop and validate a suite of baseline methods on our dataset, thereby filling a crucial missing puzzle piece in this interdisciplinary research. By releasing this dataset and its baselines, we aim to equip astrophysicists with new analytical tools while attracting data scientists to advance exoplanet direct imaging, thus catalyzing major interdisciplinary breakthroughs.

\end{abstract}

\section{Introduction}
 
\begin{wrapfigure}{r}{0.4\textwidth}
\vspace{-15pt}
\begin{center}
\includegraphics[width=0.4\textwidth]{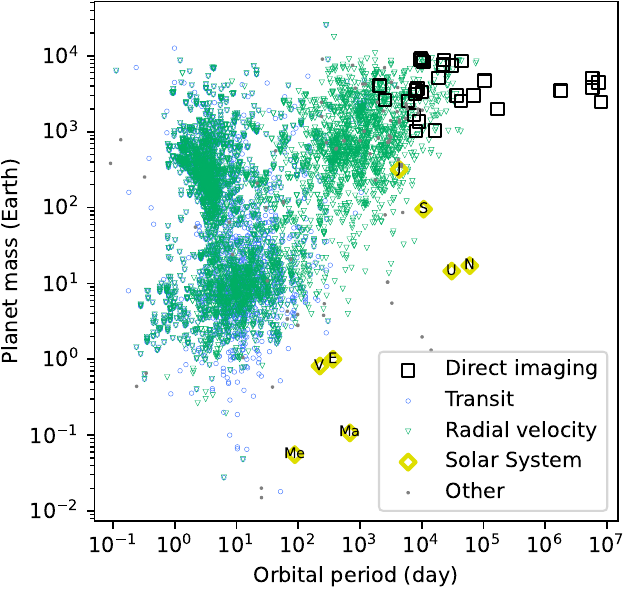}
\end{center}
\vspace{-15pt}
\caption{Mass-period distribution of known exoplanets does not reproduce the Solar System.  
Albeit with limited detections now, direct imaging probes a complementary parameter space (i.e., long-period) in exoplanet distribution, and it would uniquely reach exo-Earths in the 2030s \cite{chauvin24, stark24}.}\label{fig-distribution}
\vspace{-15pt}
\end{wrapfigure}

Since the 1995 discovery of the first exoplanet orbiting a Sun-like star \cite{mayor95}, the confirmation and diversity of the over 5800 exoplanets to date\footnote{NASA exoplanet archive (\url{https://exoplanetarchive.ipac.caltech.edu}), retrieved 2025 May 12.\label{fn-mass-period}} have revolutionized our understanding of the formation and evolution of planetary systems (e.g., \cite{lin96, dong16, owen17, bae19}). Despite these advancements,  
resemble the Solar System, let alone Earth, see Figure~\ref{fig-distribution}. These discrepancies are not just due to the uniqueness of the Solar System, but also the sensitivity limits in telescope instrumentation \cite{currie23}. Technical developments scheduled in the next 10 years would allow direct imaging to uniquely detect and characterize the first Earth-like planets (i.e., exo-Earths: \cite{chauvin24, stark24})To this point, however, direct imaging has detected less than 40 exoplanets, and only a handful of them in the past decade \cite{currie23}.

\begin{figure}[htb!]
\centering
 	\includegraphics[width=0.8\textwidth]{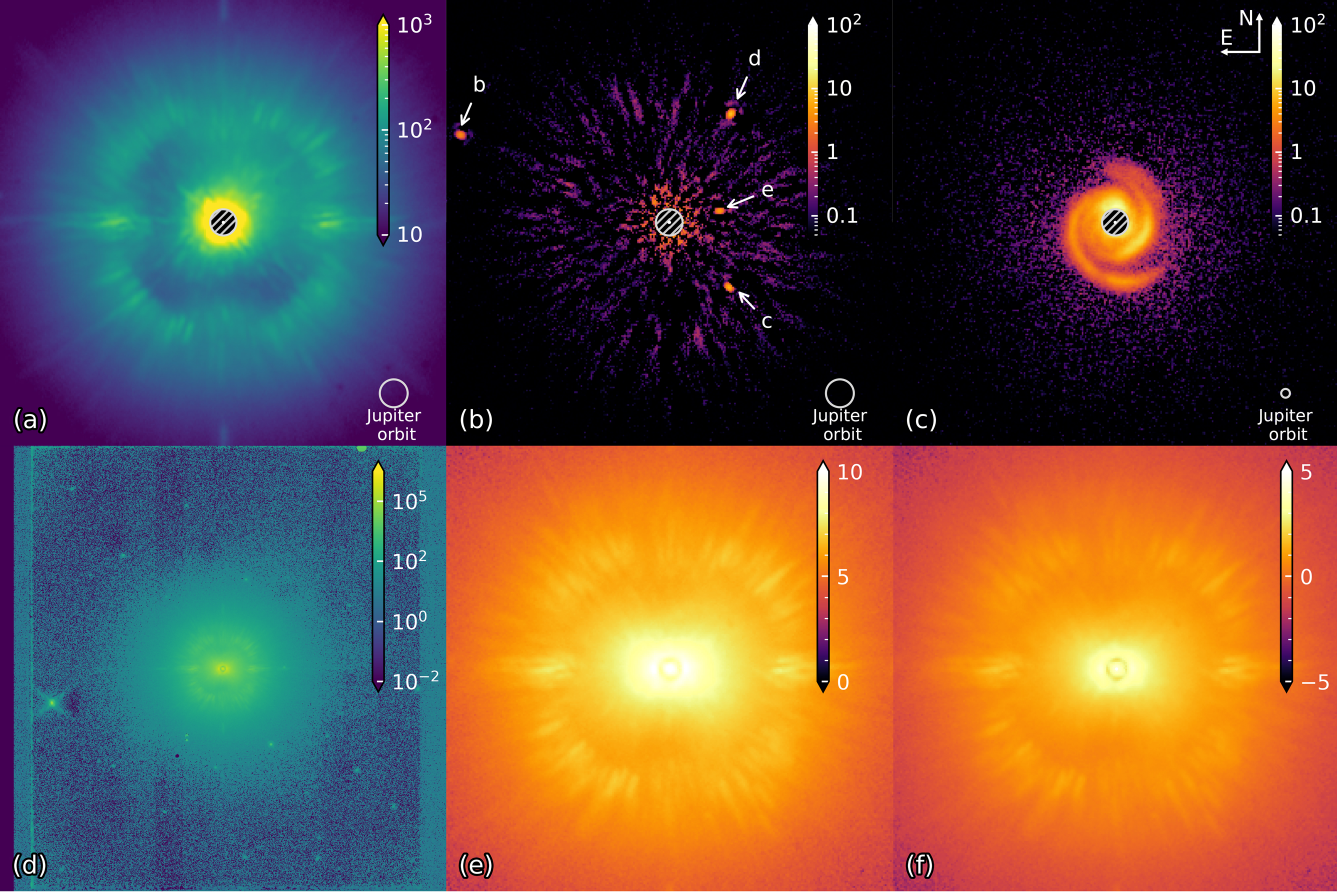}
     \caption{HCI directly images exoplanetary systems. (a) Preprocessed exposure, where star light dominates the entire field of view. (b) Originally buried in (a), four exoplanets exist around star HR~8799 in total intensity after postprocessing (e.g., \cite{wahhaj21}).
      (c) Spirals around star MWC~758 in polarized light. Note: 
      The units are detector count~s$^{-1}$~pixel$^{-1}$, and  
      central regions with 8~pixel radii (1~pixel = 12.25 mas; \cite{maire16}) are blocked by coronagraph and thus inaccessible.
      (d) Preprocessed 1024 $\times$ 1024 pixel reference image without target disk. 
      (e) Cropped to central 256 $\times$ 256 pixel area. 
      (f) Preprocessed reference data mapped to linear space.
     }
     \label{fig-image}    
 \end{figure}

In spite of its unique access to exo-Earths in the 2030s, direct imaging tackles the extreme relative faintness between the exoplanets and their host stars in visible to near-infrared light \cite{zurlo24}, i.e., high-contrast imaging (HCI). In fact, for Sun-like stars, the contrast is ${\sim}10^{-6}$ for exoplanets with several Jupiter mass (Figure~\ref{fig-image}b), or $10^{-10}$ for Earth-like ones which are not yet accessible now \cite{pueyo18}. To complement our knowledge of planetary systems, dedicated HCI surveys (e.g., Gemini Planet Imager: GPI \cite{gpi, nielsen19}, Very Large Telescope/SPHERE \cite{sphere, desidera21}, SCExAO/CHARIS \cite{charis}) have provided high-quality datasets since 2014.\footnote{The High Contrast Data Centre contains some currently public HCI datasets at \url{https://hc-dc.cnrs.fr}.\label{fn-hcdc}} However, the lack of comparable data reduction methods still limits the HCI performance \cite{bonse25, ren23}.

HCI techniques, supported by advances in both observing strategy and data reduction, have revealed exoplanetary systems even in archival datasets \cite{soummer11, bonse25}. Angular differential imaging (ADI; \cite{marois06}) exploits sky rotation during an observation to separate the static stellar point spread function (PSF) from astrophysical signals, and has proven effective in detecting compact companions like exoplanets and brown dwarfs \cite{pueyo16}. However, ADI can distort extended structures such as circumstellar disks due to self-subtraction effects \cite{milli12}. Reference differential imaging (RDI; \cite{smith84}) addresses this by using contamination-free reference stars to isolate and subtract stellar light, enabling improved recovery of extended features \cite{soummer14}. These recovered disk morphologies have not only revealed over a hundred systems \cite{benisty23}, but also hinted at embedded exoplanets \cite{dong15, bae16}, some of which are pending confirmation \cite{cugno24}. Additionally, polarimetric differential imaging (PDI) leverages polarization optics to image circumstellar disks with minimal artifacts \cite{benisty15, perrin15}, making it a powerful complement to RDI in characterizing planet-forming environments.

Exoplanet imaging with RDI requires high-quality reference star images that are free of circumstellar disk signals, yet the selection of such references has traditionally relied on manual inspection \cite{xie22, ren23, olofsson24}. Given their minimal distortion and well-characterized morphology, PDI products provide an ideal basis for automating this reference selection process. With the release of the POLARIS archive, it is now feasible to develop learnable, automated classification frameworks, reducing the need for labor-intensive labeling. Leveraging the manually curated dataset from \cite{ren23}, we extend annotations across the entire public observations from the Spectro-Polarimetric High-contrast Exoplanet REsearch (SPHERE) instrument at the Very Large Telesceope (VLT), specifically the IRDIS PDI archive—of which only $\sim$10\% had previously been labeled—resulting in a comprehensive, high-quality reference star catalog.

This constructed archive has the potential to eliminate the need for observing dedicated reference stars during telescope time—a long-standing practice in HCI \cite{wahhaj21, ren23}. Such a shift could reduce observational costs by up to $\sim50\%$, translating to approximately \$350k in savings over ten nights \cite{10.1117/12.459868}. Moreover, because circumstellar disks show stable morphology across instruments, models trained on SPHERE/IRDIS data are expected to generalize to other platforms such as GPI, CHARIS, and the upcoming Roman Space Telescope \cite{bailey23}.

\begin{figure}[htb!]
\centering
 	\includegraphics[width=\textwidth]{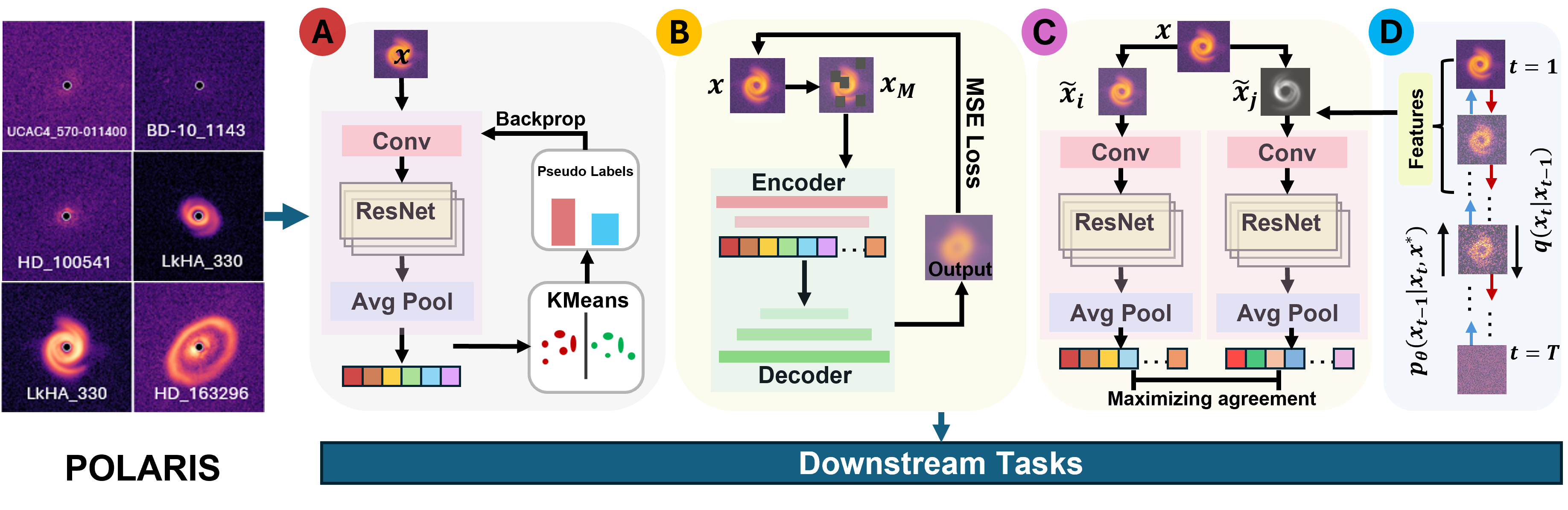}
\caption{Comparison of baselines (excluded LVLMs) and our proposed approach for representation learning on the POLARIS dataset.
(A) DeepCluster;
(B) Mask-AutoEncoder;
(C) Contrastive learning framework (SimCLR);
(D) Proposed Diff-SimCLR, which enhance latent feature representations through Diffusion for contrastive learning.}
     \label{overview}    
\end{figure}

To support this vision, we introduce a benchmark for automating two core components of RDI-based total intensity reconstruction. As Figure \ref{overview} shown, we evaluate a diverse suite of baseline models—from unsupervised learning and probabilistic generative approaches to vision-language foundation models—and further propose an unsupervised generative representation learning framework that unifies these paradigms. This framework not only achieves state-of-the-art performance in classification, but also yields high-fidelity priors for downstream tasks such as background reconstruction and image enhancement. By benchmarking these methods and providing labeled RDI reference images at scale, this work lays the foundation for scalable, automated exoplanet imaging and enables rigorous comparison of deep learning approaches—an essential capability that has been largely absent from the field for decades.

\vspace{-6mm}

\section{Related Work}

\textbf{Data Repositories:} Instrument development advancements on PDI in the past decade has greatly revolutionized our understanding of circumstellar environments (e.g., \cite{tamura16, avenhaus18, esposito18, rich22}), since PDI can uniquely access polarized light which trace dust scatterers in exoplanetary systems with the least bias \cite{monnier19}. However, with different instrument setups and data reduction strategies, the public PDI \Qphi\ files have been mostly limited to individual scientific publications (e.g., \cite{rich22, ren23}). While the HC-DC have the postprocessed results for a number of HCI surveys,\textsuperscript{\ref{fn-hcdc}} the raw exposures -- as well as the preprocessed files for later postprocessing reduction -- are not systematically accessible beyond the astronomy community.

\textbf{ML Benchmarks:}
Current methods for the recovery of circumstellar objects in HCI typically rely on the availability of corresponding reference star images to enable the extraction of exoplanetary signals while minimizing the impact of self-subtraction artifacts \cite{pueyo18}. The most widely adopted approach involves eigenvalue decomposition and subspace projection, where information from reference images is used to model the stellar PSF. This modeled background is then projected onto the target image, facilitating stellar light subtraction through residual differencing \cite{soummer14}. A prominent example is the Karhunen–Loève Image Projection (KLIP) algorithm \cite{soummer12}, which has become a standard technique in current exoplanet imaging surveys \cite{nielsen19, langlois21}. Recent advancements, driven by parallel computing and iterative optimization techniques, have yielded more sophisticated approaches that outperform classical methods. These include iterative frameworks such as 4S \cite{bonse25}, iterative PCA \cite{juillard24}, and matrix decomposition-based methods like MAYONNAISE \cite{pairet21}, non-negative matrix factorization \cite{ren18}, and REXPACO \cite{flasseur21}. While these methods have demonstrated improved performance in specific systems, their validation has generally been limited in scope (i.e., only several systems). This narrow validation hampers both their generalizability and scalability when applied across broader, heterogeneous datasets. A further challenge lies in the classification and selection of reference and target images, which remains heavily reliant on heuristic or empirical strategies. These manual or semi-automated approaches are often computationally expensive and labor-intensive, especially when scaled to large archives or survey programs involving many stellar systems \cite{Xie2022}.
\section{Overview of POLARIS Dataset}\label{sec-data}
\subsection{Data Collection}
POLARIS is based on a decade of polarimetric observations obtained with VLT/SPHERE from 2014 to 2024.\footnote{Available from European Southern Observatory (ESO) Science Archive Facitlity at \url{http://archive.eso.org/wdb/wdb/eso/sphere/form}. The observations are normally public after 12 months of proprietary period.} Specifically, we retrieved the entire public observations using the SPHERE's IRDIS instrument in polarized light. To prepare the raw observational data for analysis, we follow \cite{ren23} to adjust\footnote{GitHub repo: \url{https://github.com/seawander/IRDAP}, which is adjusted from the original one at \url{https://github.com/robvanholstein/IRDAP}.} the \texttt{IRDAP} \cite{irdap1, irdap2} data reduction pipeline to both uniformly preprocess the datasets and obtain PDI-postprocessed products. By manually inspecting the preprocessed and \Qphi\ files, we removed bad exposures (e.g., raw files, calibration files, star centering files), and reran \texttt{IRDAP} to ensure POLARIS data quality.  The final POLARIS \Qphi\ files from PDI are particularly effective at revealing light scattered by dusty circumstellar disks (e.g., \cite{monnier19}), and thus a non-detection of such signals in a \Qphi\ file can identify the corresponding original exposures as potential reference images.

Our POLARIS dataset also contains individual \texttt{IRDAP}-preprocessed exposure sequences, with each sequence operating HCI for a chosen star in a $1$--$2$~hour observation block. In a sequence, the polarization optical component normally cycles through Stokes \{Q$^+$, Q$^-$, U$^+$, U$^-$\} exposures to enable PDI \cite{irdis}, totaling $4n$ images ($n\in\mathbb{Z}$) per sequence when it is not interrupted. Calibration exposures, which are not included in POLARIS, are taken during an observation for \texttt{IRDAP} to remove bad pixels, center the images, and remove sky thermal background \cite{irdap1}. The preprocessed exposures in one sequence are used by \texttt{IRDAP} to produce one \Qphi\ file. In POLARIS, there are currently 921 polarized \Qphi\ files (for labeling), as well as the corresponding $75,910$ preprocessed files (for data imputation).\footnote{The normal VLT/SPHERE operation and ongoing upgrade \cite{mazoyer24} ensure continuous dataset expansion.\label{fn-cont}}  Among the \Qphi\ files, 96 are already labeled as either targets or references \cite{ren23}. A target corresponds to a planetary system exhibiting a prominent circumstellar structure (e.g., spirals, rings), whereas a reference has a non-detection of such structures and it serves as the background context (i.e., star-only signals).  
A sequence of preprocessed images would be classified as reference exposures once their corresponding \Qphi\ image does not host circumstellar structure. Both the \Qphi\ and the preprocessed files are stored in \texttt{.fits} format \cite{pence10} following astronomy standards.

\subsection{Data Preprataion}

One \texttt{IRDAP}-preprocessed file consists of time-series images with shape $(n, 1024, 1024)$, where $n$ denotes the total number of files in an observation sequence (Figure~\ref{fig-image}d).  For both classification and imputation tasks, we crop and normalize the central $256 \times 256$ pixel region (Figure~\ref{fig-image}e). Pixel values represent light intensity received by the HCI detector, ranging approximately from $-10^2$ to $10^5$ counts~s$^{-1}$~pixel$^{-1}$, where negative values are non-physical due to detector or observation imperfection. To stabilize the dynamic range, we apply a logarithmic transformation after setting negative values to zero. The resulting images are then linearly rescaled to the range $[-4, 4]$ (Figure~\ref{fig-image}f). To support research on our POLARIS representation learning, we create Single-frame polarimetric images (\texttt{.fits}, $256 \times 256$) are normalized to the range $[0, 1]$, saved in \texttt{.jpeg} format, and stored as NumPy arrays. Preprocessed exposure sequences are stored as \texttt{.fits} files with shape $(4n, 256, 256)$, where $4n$ corresponds to multi-cycle temporal exposures, and are also saved as NumPy arrays. Both data types share matching filenames (system name and observation date) to enable alignment of classification results with their corresponding exposure sequences.

The dataset introduced in this work, \textbf{POLARIS}, is publicly available on \href{https://zenodo.org/records/15427454}{Zenodo}.
It comprises: (i) \textbf{96 labeled PDI-postprocessed polarimetric images} ($1024 \times 1024$ pixels), annotated as either \emph{target} or \emph{reference}, archived at approximately 30 MB; (ii) \textbf{813 unlabeled PDI-postprocessed images}, derived from \emph{preprocessed total intensity exposures} from 2014–2023, each annotated with vegetation indices and land-use metadata, totaling around 400 MB; the 2024 data will be included in the next version, bringing the total to 921 images; and (iii) the corresponding preprocessed exposure sequences ($4n \times 1024 \times 1024$) from 2014–2024, where $n$ is the number of exposures per sample, exceeding 200 GB in total.
All POLARIS data are provided as compressed \texttt{.zip} archives, with preprocessed exposures hosted via a Dropbox link. All experiments in this paper use \textbf{versions 1.0–2.0} of the dataset. Future versions will be versioned and archived on \href{https://zenodo.org/records/15427454}{Zenodo} for reproducibility.

\section{Tasks and Baseline Experiments}

\subsection{Unsupervised Learning on POLARIS}

\subsubsection{Baseline Frameworks} 
\label{sec:baseline_frams}

We evaluate three baseline methods commonly used in unsupervised feature learning for the POLARIS dataset, aimed at supporting representation learning in the latent space, along with one proposed method. The baselines include two self-supervised learning frameworks—Masked Autoencoder (MAE)~\cite{he2021maskedautoencodersscalablevision} and DeepCluster~\cite{caron2019deepclusteringunsupervisedlearning, ren2022deepclusteringcomprehensivesurvey}—as well as an unsupervised contrastive learning approach, SimCLR~\cite{chen2020simpleframeworkcontrastivelearning, khosla2021supervisedcontrastivelearning}. Our proposed method, Diff-SimCLR, extends SimCLR by incorporating a diffusion-based module to enhance latent feature representations. As the models are designed to learn informative representations for subsequent classification tasks, a 32-dimensional feature vector is selected as the output representation. This dimensionality reflects a balance between sufficient representational capacity and computational feasibility for downstream tasks, while also mitigating model complexity due to the limited size of the labeled dataset. Note that, we tune the hyperparameters by grid search for all models. 

\textbf{MAE:} The framework contains a vision transformer (ViT)~\cite{DBLP:journals/corr/abs-2010-11929} encoder on unmasked patches and a MAE decoder contains visible patches and mask tokens with positional embeddings \cite{he2021maskedautoencodersscalablevision}. An optimal masking ratio of 20\% is applied to the input image, with visible patch sizes of (16,16). The model learns to infer missing regions and is trained using mean squared error loss between the reconstructed and original images. The modified network consists of a convolutional autoencoder that progressively reduces the spatial dimensions of 256$\times$256 grayscale images, ultimately encoding each input into a compact 32-dimensional latent representation, which is then decoded for reconstruction under incomplete input conditions. The network is trained for 150 epochs with an optimized learning rate of $1e^{-4}$ and batch size of 32.

\textbf{DeepCluster:} The deep clustering framework involves passing data through a feature learning network, using the learned features for clustering, and generating corresponding pseudo-labels for self-supervised learning via stochastic gradient descent (SGD) backpropagation \cite{caron2019deepclusteringunsupervisedlearning}. Instead of relying on a pre-trained convolutional network, we apply a residual network, consistent with the approach used in \texttt{SimCLR}, as it aligns well with the structure of POLARIS. This approach alternates between clustering image descriptors and updating the convolutional network's weights by predicting cluster assignments using {k-means}. The Deep Clustering framework has been refined to Python 3.11, with the network trained for 100 epochs, a learning rate of $1e^{-2}$, a batch size of 16, and k-means clustering configured with 2 clusters.

\textbf{SimCLR:} Simple framework for Contrastive Learning of visual Representations (SimCLR) model leverages the idea of contrastive learning to learn feature representations via maximizing the agreement between different augmented view of data~\cite{khosla2021supervisedcontrastivelearning}. As an augmentation of image $\tilde{x}_i$ will pass through a backbone residual network $h_i = \text{ResNet}(\tilde{x}_i)$ with output dimension 512. The final latent representation feature is the result of backbone through a multilayer perception, $z_i = \text{MLP}(h_i)$, and same for its paired augmentation $z_j = \text{MLP}(h_j)$. The model is optimized through the {NT-tent} Loss, $\mathcal{L}(z_i, z_j)$. The model is trained with 200 epochs with a learning rate of $1e^{-3}$ and batch size of 32. As meeting the agreement, the 32-dimensional feature representation $z$ will be extracted. 

\textbf{Large Vision-Language Models:} For POLARIS image classification, we design a zero-shot prompt template and instruct the LVLM to act in capacity of an exoplanet astronomer. Figure~\ref{llm_prompt} (see Appendix) shows an example prompt designed for an image in POLARIS dataset. Our expert-designed prompt consists of two parts: (i) the general prompt which introduces the task scenario and (ii) dataset description which describes the characteristics of the target and reference images we want to focus on. Thus, this designed prompt provides LVLM with the general goal and the classification task. Then we use the proposed prompt $\mathcal{P}$ to query LVLM to get the classification of the image. For an image $x_i$, the process can be formally defined as
\begin{align}
    c_i = \text{LVLM}(\mathcal{P}, x_i),
\end{align}
where $c_i \in [\text{target}, \text{reference}]$ denotes the predicted image type of $x_i$. 
We also compare the capabilities of 7 different LVLMs in analyzing our POLARIS data. For OpenAI GPT models, we access the GPT-4o and GPT-4.1 via the OpenAI API and set \texttt{temperature} to 0. For Gemini-2.0-Flash, we utilize the Google Vertex AI Cloud API and set the \texttt{temperature} to 1. In addition, we use four open-source models, i.e., Llama-3.2-11B (i.e., Llama-3.2-11B-Vision-Instruct), Llama-3.2-90B (i.e., Llama-3.2-90B-Vision-Instruct), DeepSeek-VL2-Tiny, and DeepSeek-VL2-Small and all these four models are set with a \texttt{temperature} of 0. 
 
\subsubsection{The Proposed Baseline: Latent-Enhanced Contrastive Learning (Diff-SimCLR)}
\label{sec:baseline_frams_proposed}

Recent advancements in generative models, particularly Diffusion models, have shown promising potential in enhancing representation learning in many domains \cite{10485553, Liu_2024, diffusionbasedzeroshotmedical}. Contrastive learning enables models to learn representations invariant to image augmentations \cite{zhang2022rethinking}, but these representations may still lack the compactness required to capture subtle inter-class differences. To address this, we propose enriching contrastive features with latent information extracted from a conditional denoising diffusion probabilistic model (DDPM) \cite{ddpm}, which further improves feature representation and enhances model performance.

We start with an input image \(x\) and apply two different random augmentations to create a pair of modified views, \(\tilde{x}_1\) and \(\tilde{x}_2\). The goal of our method is to learn feature representations that are consistent between these augmentations while still preserving the ability to distinguish between different classes. Each augmented image is processed by a modified ResNet backbone $f_{\text{ResNet}}$ to extract feature embeddings, denoted as $h_i = f_{\text{ResNet}}(\tilde{x}_i) \in \mathbb{R}^k, \quad i = 1, 2$. Concurrently, we extract a configurable prior from the Diffusion model by collecting the last $\Delta_t$ latent states. Let $x_t$ be the noisy version of $x$ at timestep $t$ in the diffusion process, with $x_0 = x$. The prior trajectory is defined as:
\begin{equation}
    p = [x_0, x_1, \dots, x_{\Delta_t}] \in \mathbb{R}^{(\Delta_t + 1) \times d}
\end{equation}
where we choose $\Delta_t = 8$ to balance informativeness with computational cost. The prior sequence is encoded using the same ResNet backbone: $h_p = f_{\text{ResNet}}(p) \in \mathbb{R}^k$. The two latent features are fused by concatenation and projected through a shared head $g: \mathbb{R}^{2k} \rightarrow \mathbb{R}^m$:
$z_i = g([h_i \| h_p]) \in \mathbb{R}^m$,
where $\|$ denotes vector concatenation. The output $z_i$ is used for contrastive learning.

The Diffusion model itself operates by progressively adding noise to the input image over time using a forward process: $q(x_t \mid x_{t-1}) = \mathcal{N}(x_t; \sqrt{1 - \beta_t} \, x_{t-1}, \beta_t I)$, where \(\{\beta_t\}_{t=1}^T\) is a fixed noise schedule. During inference, the model reverses this process using a denoising step that is conditioned on a noisy reference image: $p_\theta(x_{t-1} \mid x_t, x^*) = \mathcal{N}(x_{t-1}; \mu_\theta(x_t, x^*, t), \sigma_t^2 I)$, where \(x^*\) is a corrupted version of the input. This conditional denoising helps the model preserve structural information during generation, improving the quality of the learned priors.

We train the DDPM for 300 epochs with a learning rate of $1e^{-3}$ and batch size of 16. After convergence, we fix the DDPM parameters and train the contrastive model for 200 epochs with the same learning rate and a batch size of 32. The model is optimized using the InfoNCE loss on the paired embeddings $(z_1, z_2)$.

\vspace{-4mm}
\begin{table*}[htpb!]
\centering
\footnotesize	
\setlength\tabcolsep{2pt}
\caption{Comparing classification accuracy from different LVLMs.\label{lvlm_res}}
\resizebox{0.99\columnwidth}{!}{
\begin{tabular}{l c c c c c c c}
\toprule 
\textbf{Data} & \textbf{GPT-4o} & \textbf{GPT-4.1} & \textbf{Llama-3.2-11B}& \textbf{Llama-3.2-90B}& \textbf{Gemini-2.0-Flash} & \textbf{DeepSeek-VL2-Tiny}& \textbf{DeepSeek-VL2-Small} \\ 
\midrule
POLARIS & 67.71 & 75.00 & 48.96 & 52.08 & {\bf 75.21} & 49.12 & 50.00\\
\bottomrule
\end{tabular}}
\end{table*}
\subsection{Classification on POLARIS: Evaluating Downstream Task Performance}
\label{downstream_task_res}

\textbf{Downstream Tasks:} To evaluate the quality of the learned latent features, we extract a representative result—specifically, a 32-dimensional feature vector for each of the 96 labeled images—using the aforementioned frameworks trained on unlabeled data. Four \textit{supervised downstream classification} tasks are applied : linear Support Vector Classifier (SVC), kernel Support Vector Machine (SVM), Random Forest, and Multi-Layer Perceptron Classifier (MLPClassifier). Regression tasks are excluded due to overfitting risk, as indicated by the high events-per-variable ratio \cite{vanSmeden2016}. Hyperparameters for all classifiers are optimized within a searching region. 
A 10-fold Stratified Cross-Validation (CV) procedure is applied, where hyperparameters are fine-tuned within each fold using a 5-fold grid search. The classifier is trained on the training data of each fold and evaluated on the test data. The final performance is reported as the mean accuracy across all folds. 
Three \textit{unsupervised downstream tasks} are employed to evaluate the suitability of the learned features for classification: K-Nearest Neighbors (KNN), Gaussian Mixture Model (GMM), and Spectral Clustering. KNN is applied with 2 clusters and 30 iterations. GMM uses an isotropic covariance structure to mitigate overfitting. Spectral Clustering includes a 5-fold grid search, varying the number of neighbors $n \in \{3, 5, 7, 10\}$ to examine local connectivity, and tests both \textit{k}-means and discretization methods for label assignment. Cluster labels are aligned to ground truth using the Hungarian algorithm for optimal matching. All evaluations are conducted with 10-fold CV and a fixed random seed, and we report the mean accuracy across folds. For further
details, please refer to Appendix.

\begin{figure}
    \centering
\includegraphics[width=\linewidth]{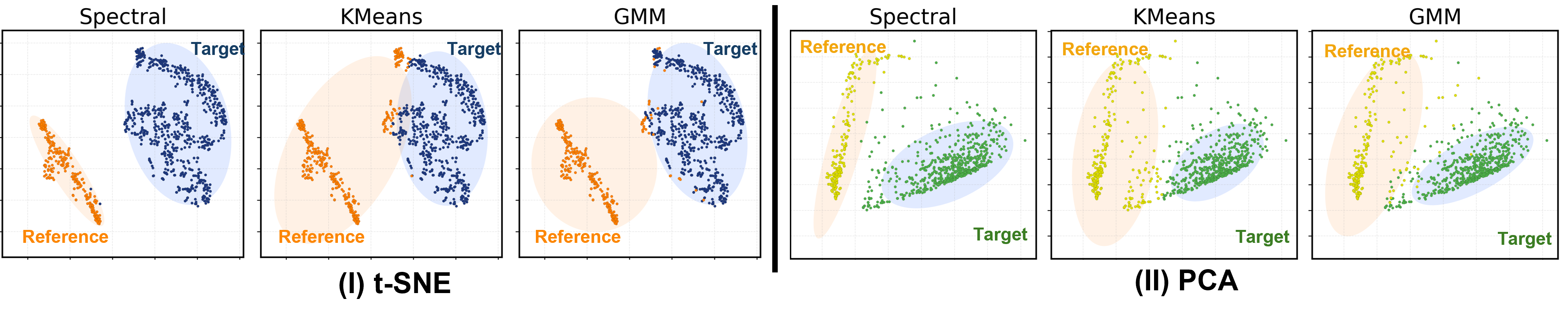}
    \caption{
    \textbf{(I):} Unsupervised clustering results of features extracted by Diff-SimCLR from 813 unlabeled polarized images, reduced to two dimensions via t-SNE ($32 \rightarrow 2$), using three downstream clustering methods: Spectral Clustering, K-Means, and Gaussian Mixture Modeling. Final label assignments for the two clusters are performed using the Hungarian algorithm to maximize agreement with the labeled images. 
    \textbf{(II):} Corresponding clustering results visualized with PCA for dimensionality reduction ($32 \rightarrow 2$).The clustering methods exhibit general agreement, with Spectral Clustering producing the most distinct separation, particularly for reference images. }
    
\label{fig:unsupervised_clustering}
\end{figure}

\textbf{Disk Classification:} Table~\ref{lvlm_res} shows the classification results on our POLARIS dataset. Our observations are: (i) Compared to other LVLMs, Gemini-2.0-Flash achieves the highest performance with yielding 31.92\% relative improvement on average, which can be interpreted as a significant improvement. Specifically, compared to three open-source LVLMs (i.e., Llama-3.2-11B, Llama-3.2-90B, DeepSeek-VL2-Small), Gemini-2.0-Flash achieves on average 49.48\% relative improvement and (ii) Both GPT-4o and GPT-4.1 deliver highly competitive results which achieve on average 41.82\% relative improvement over open-source LVLMs. These results decisively demonstrate that the effectiveness and potential of LVLMs in analyzing future large-scale polarimetric
images. 
Table~\ref{res_3} reports the downstream classification accuracy on the 32-dimensional feature representations extracted from POLARIS using four representative classification models. The first four columns correspond to supervised learning methods. Among these, the proposed latent-enhanced contrastive learning approach (Diff-SimCLR) consistently outperforms the alternatives across all classifiers, achieving the highest accuracy of 93.00\% with the SVC, as also reflected in Table~\ref{res_2}. The unsupervised clustering of Diff-SimCLR features in Figure \ref{fig:unsupervised_clustering} aligns with the quantitative results, with both t-SNE and PCA visualizations highlighting the effectiveness and separability of the learned representations. This observation indicates that the features learned by our proposed Diff-SimCLR effectively capture object types and structural characteristics in \Qphi\ polarized HCI images, supporting both robust and interpretable classification. The last three columns represent unsupervised learning methods. While Diff-SimCLR features demonstrate strong and stable clustering performance, they generally underperform relative to supervised approaches which highlight the inherent challenge of label-free discrimination in this domain.

\begin{wrapfigure}{r}{0.45\textwidth}
\vspace{-15pt}
\begin{center}
\includegraphics[width=0.45\textwidth]{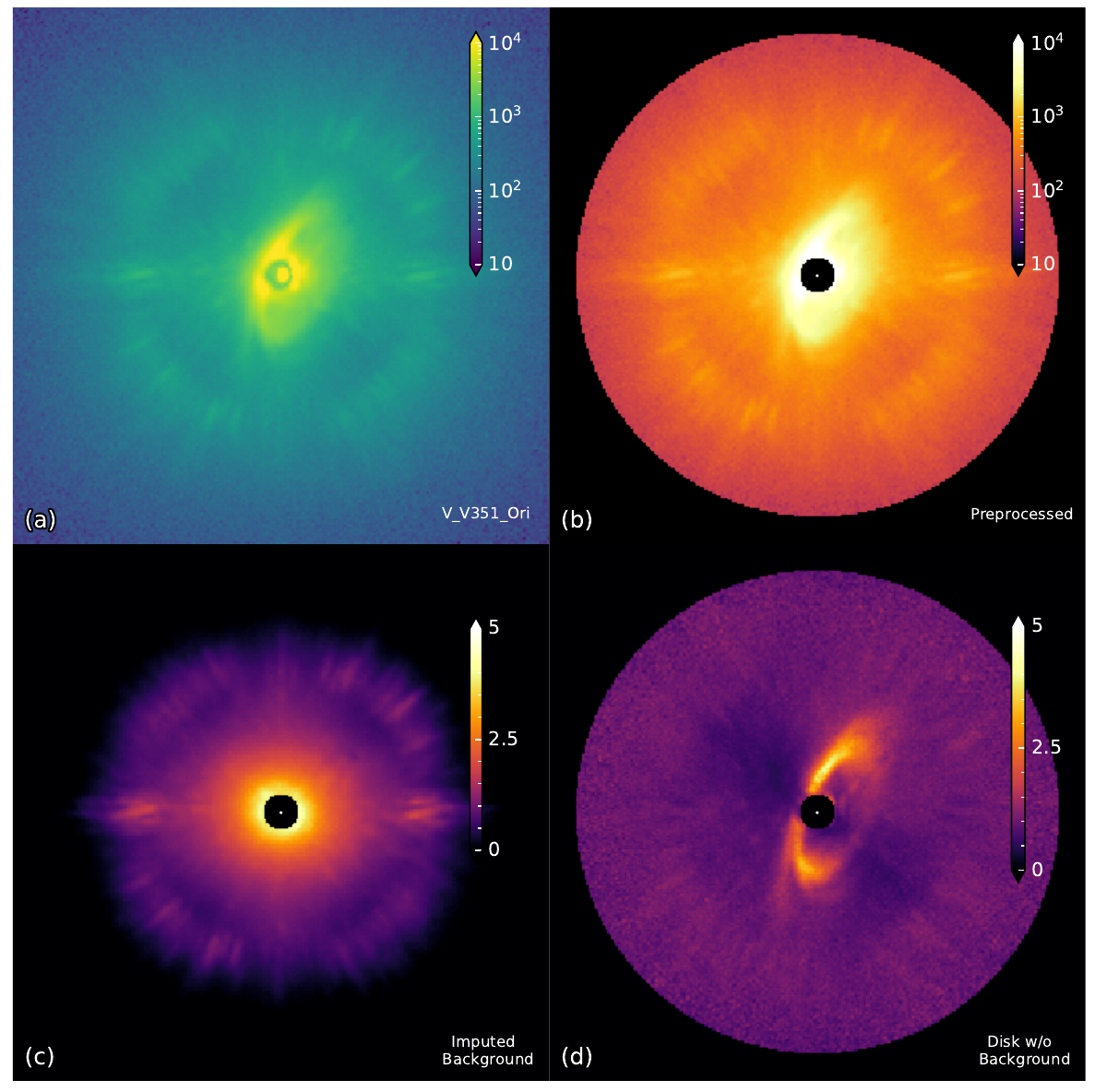}
\end{center}
\vspace{-15pt}
\caption{VAE-based reconstruction of circumstellar disk. (a) Original image with background star. (b) Preprocessed. (c) VAE-predicted background. (d) Disk after background subtraction.}
\label{vae_disk}
\vspace{-15pt}
\end{wrapfigure}

\textbf{Preliminary Verification on Disk Reconstruction:} Spectral clustering is selected to assign label information to the unlabeled PDI-postprocessed polarimetric images, based on its superior accuracy on a reference set of labeled images (see Table~\ref{res_3}), representing the most stringent evaluation criterion. The clustering result is obtained using a nearest neighbors parameter of 7 and the discretized label assignment method, which offers more stable and deterministic performance compared to alternative approaches. These settings are identified as optimal through the CV pipeline outlined in Section~\ref{AppendixA2}. Classification outputs include reference star system names and observation dates matched to corresponding preprocessed exposure sequences, which serve as input for a preliminary background imputation task assessing the viability of a probabilistic modeling approach.
A total of 206 images are assigned to reference clusters, and their corresponding exposures are used to train a variational autoencoder (VAE) for background reconstruction. Sequential images are cropped to a $256 \times 256$ central region with the central 8-pixel radius excluded—matching the coronagraphic occulter in IRDIS to avoid saturation—and log-transformed, with up to four frames per exposure fed into the model (see Figure~\ref{fig-image}). The central region of radius 80 pixels is masked during training, enabling the VAE to learn background structure surrounding this area. The encoder employs convolutional layers and max pooling to reduce inputs to a $32 \times 32$ latent representation, while the decoder reconstructs images via transposed convolutions back to full resolution. The composite loss function incorporates masked reconstruction error, Kullback-Leibler (KL) divergence regularization, boundary consistency, and pattern preservation through directional kernels and alongside normalization alignment to better capture intensity statistics. After training, masked exposures from target images are processed through the model to infer central background star PSF information, which is then subtracted from the target images to isolate the circumstellar disk signal.
The result is showed in Figure~\ref{vae_disk}, background pattern such as Airy disk is well imputed by VAE model that the simulated light track aligned the original central background star PSF information. The target disk explicitly appeared when the star background noise, to some extent, is removed. With VAE model's help, traditional star PSF background clean-up work, in which the astronomers manually fitting suitable star systems, is replaced by this powerful \textit{AI + Exoplanetary System} tool.

\vspace{-3mm}
\begin{table*}[htpb!]
\centering
\footnotesize
\setlength\tabcolsep{2pt}
\caption{Performance comparison among different machine learning classifiers.\label{res_3}}

\definecolor{lightpurple}{RGB}{238, 230, 255}
\definecolor{lightgreen}{RGB}{230, 255, 208}

\begin{tabular}{l c c c c c c c}
\toprule 
\textbf{Model} & \textbf{SVC} & \textbf{Random Forest} & \textbf{MLPClassifier} & \textbf{SVM} & \textbf{KNN} & \textbf{GMM} & \textbf{Spectral}\\ 
\midrule
Maskencoder & 
\cellcolor{white}80.33 & \cellcolor{white}77.44 & \cellcolor{white}82.29 & \cellcolor{white}85.00 & 
\cellcolor{white}73.78 & \cellcolor{white}74.00 & \cellcolor{white}77.00\\

SimCLR & 
\cellcolor{white}84.78 & \cellcolor{white}84.33 & \cellcolor{white}82.00 & \cellcolor{white}86.46 & 
\cellcolor{white}73.89 & \cellcolor{white}71.11 & \cellcolor{white}\textbf{77.78}\\

DeepCluster & 
\cellcolor{white}67.67 & \cellcolor{white}74.00 & \cellcolor{white}70.83 & \cellcolor{white}69.67 & 
\cellcolor{white}70.67 & \cellcolor{white}72.00 & \cellcolor{white}74.89\\
\hline
\textbf{Diff-SimCLR} & 
\cellcolor{white}\textbf{93.00} & \cellcolor{white}\textbf{89.67} & \cellcolor{white}\textbf{92.71} & \cellcolor{white}\textbf{89.56} & 
\cellcolor{white}\textbf{75.00} & \cellcolor{white}\textbf{74.22} & \cellcolor{white}77.33\\
\bottomrule
\end{tabular}
\end{table*}
\vspace{-3mm}

\section{Broader Impact}

We have labeled public IRDIS polarized archive here, and existing and upcompoing observations with existing instruments can directly benefit from our work.  In fact, 
SPHERE has three instruments \cite{sphere}: 
ZIMPOL in visible light \cite{zimpol}, IFS in multiple wavelengths (${>}30$ wavelength channels/images in one exposure: \cite{ifs}), and IRDIS in the near-infrared (either polarization observations here, or total-intensity-only: \cite{irdis, xie22}). For all HCI systems (e.g., SPHERE, GPI, SCExAO), 
once a star is identified by any instrument as a target, it can be directly labeled as targets for all instruments. 
Future telescopes would directly benefit from the exploration in this work. First, the 2.4 meter Roman Space Telescope in 
${\sim}$2027: its Coronagraph Instrument \cite{bailey23} requires dedicated reference star vetting for exoplanet imaging. 
Second, the ground-based 40-meter Extremely Large Telescope (ELT) -- which is over 20 times the collection area of VLT -- by 2030 has unprecedented sensitivity (down to Earth-sized exoplanets: \cite{quanz15}).

RDI is more observationally economic for ELT, since ADI requires sky rotation and thus a large integration number, and RDI would thus directly benefit from the explorations here. Third, NASA will launch its next space-based flagship mission, the 6 meter Habitable Worlds Observatory (HWO) in ${\sim}2035$ that will image and characterize exo-Earths \cite{stark24}. 

Furthermore, we benchmark several generative models, including a VAE, to evaluate their effectiveness. Notably, we show that realistic stellar backgrounds can be synthesized directly from target images using models trained exclusively on RDI data, demonstrating strong transferability. This approach enables circumstellar disk reconstruction without requiring manually paired reference images, offering a scalable alternative to conventional reference selection in exoplanet imaging. 
Apart from above-mentioned missions, our work here suggests that AI methods for other tasks (e.g., wavefront control) might reduce human efforts, and thus ensure mission success and maximize scientific impact. The ability to automatically label targets across instruments significantly reduces the reliance on manually curated catalogs. Our studies support the development of generalizable learning algorithms capable of integrating heterogeneous data types (e.g., polarization, spectral channels) and align with ongoing efforts in AI and translational science. In addition, the initiative on \textit{AI + Exoplanetary System} opens up avenues for rigorous modeling of astrophysical signals under noisy and high-uncertainty conditions.
\section{Limitations}
The 96 manually labeled systems in \cite{ren23} are the brightest circumstellar objects -- protoplanetary disks -- where active giant planet formation is ongoing \citep{ikoma25}. However, circumstellar disks from protoplanetary disks dissipate to debris disks \cite{kral18}, and disk would be significantly fainter given the mass reduction of ${>}10^3$. 
Indeed, POLARIS contains debris disk observations (e.g., \cite{adam21}; Appendix Figure~\ref{fig-app-gallery}), and they could be identified as false negatives for targets. While the impact including debris disk exposures in the references in recovering stellar-light-only signals for protoplanetary disk targets using RDI might be small, it prevents a proper detection and characterization of faint debris disks. We assume that a non-detection of circumstellar objects in polarized light is equivalent to their non-existence in total intensity. While this is true for circumstellar disks, it is not for other objects such as exoplanets. Although exoplanets have been rarely imaged beyond $30$~au (Figure~\ref{fig-distribution}), they would populate the $3$--$10$~au region \cite{fulton21} for future HCI. This is an expected challenge for labeling using future observations, and it would potentially require point source identification for future methods. In fact, once a point source is identified, we can mask it out and use multiple masked images to self-impute themselves (e.g., ADI with missing data: \cite{ren20}).

\section{Conclusion}
We introduce the POLARIS dataset—a large-scale, high-quality benchmark for polarimetric representation learning in exoplanet imaging. Derived from a decade of SPHERE/IRDIS polarized observations, POLARIS provides both labeled and unlabeled data that enable scalable learning of reference-star classification and circumstellar disk detection. We systematically evaluate a suite of statistical, generative, and LVLMs, establishing baseline performance and releasing reproducible code and evaluation protocols. Motivated by the growing utility of generative AI, we further propose an unsupervised generative representation learning framework, Diff-SimCLR, which achieves state-of-the-art accuracy in both supervised and unsupervised settings. To our knowledge, this is the first ML benchmark designed specifically for exoplanet imaging. By bridging astronomy and machine learning through this open benchmark, we aim to accelerate methodological innovation and enable more efficient, data-driven discovery in future HCI surveys.

\clearpage
\bibliographystyle{plain}
\bibliography{references}


\clearpage

\appendix

\section{Technical Appendices and Supplementary Material}

\subsection{Supplementary Details on POLARIS}
\label{AppendixA1}

\textbf{Note that} the latest release of the POLARIS dataset (version~v2) is available on \href{https://zenodo.org/records/15449056}{Zenodo} and is consistent with the accompanying Croissant metadata file. For reference, version~1.0 remains accessible via \href{https://zenodo.org/records/15427454}{this link}.

\begin{figure}[htb!]
 \vspace{2mm}
\centering
 	\includegraphics[width=\textwidth]{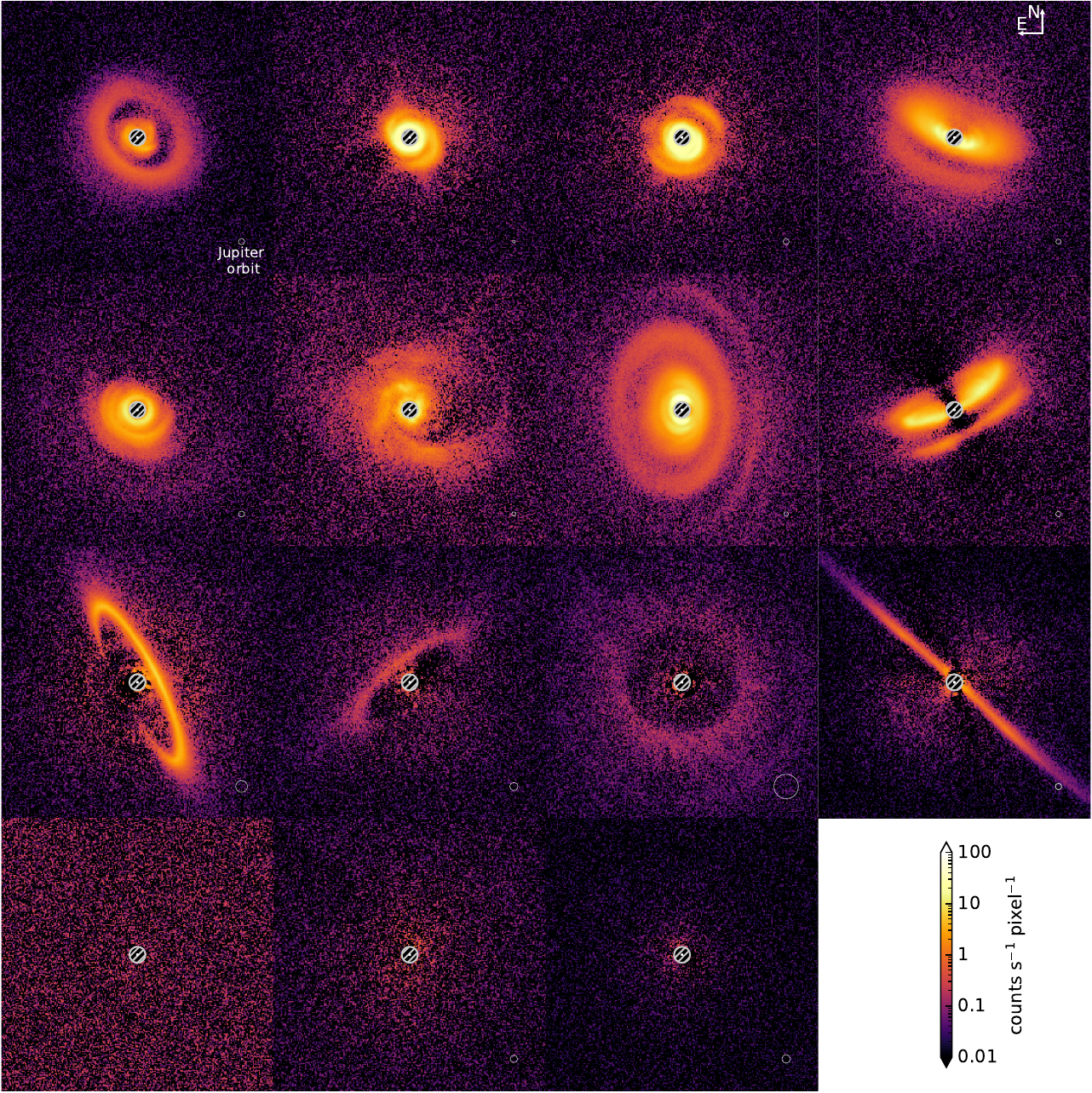}
     \caption{Sample POLARIS \Qphi\ images. 1st and 2nd row: protoplanetary disks, which are relatively bright. 3rd row: debris disks, which are relatively faint. 4th row: reference stars. Notes: (1) The panels here share the same field of view and color bar, with the central regions with 8~pixel radii blocked, and the lower right circle in each panel denotes Jupiter orbit ($5.2$~au), i.e., the setup in Figure~\ref{fig-image}. (2) The 4th column shows diverse morphology for (nearly) edge-on systems that are not included in \cite{ren23}.}
     \label{fig-app-gallery}    
 \end{figure}

\begin{figure}[htb!]
\centering
 	\includegraphics[width=\textwidth]{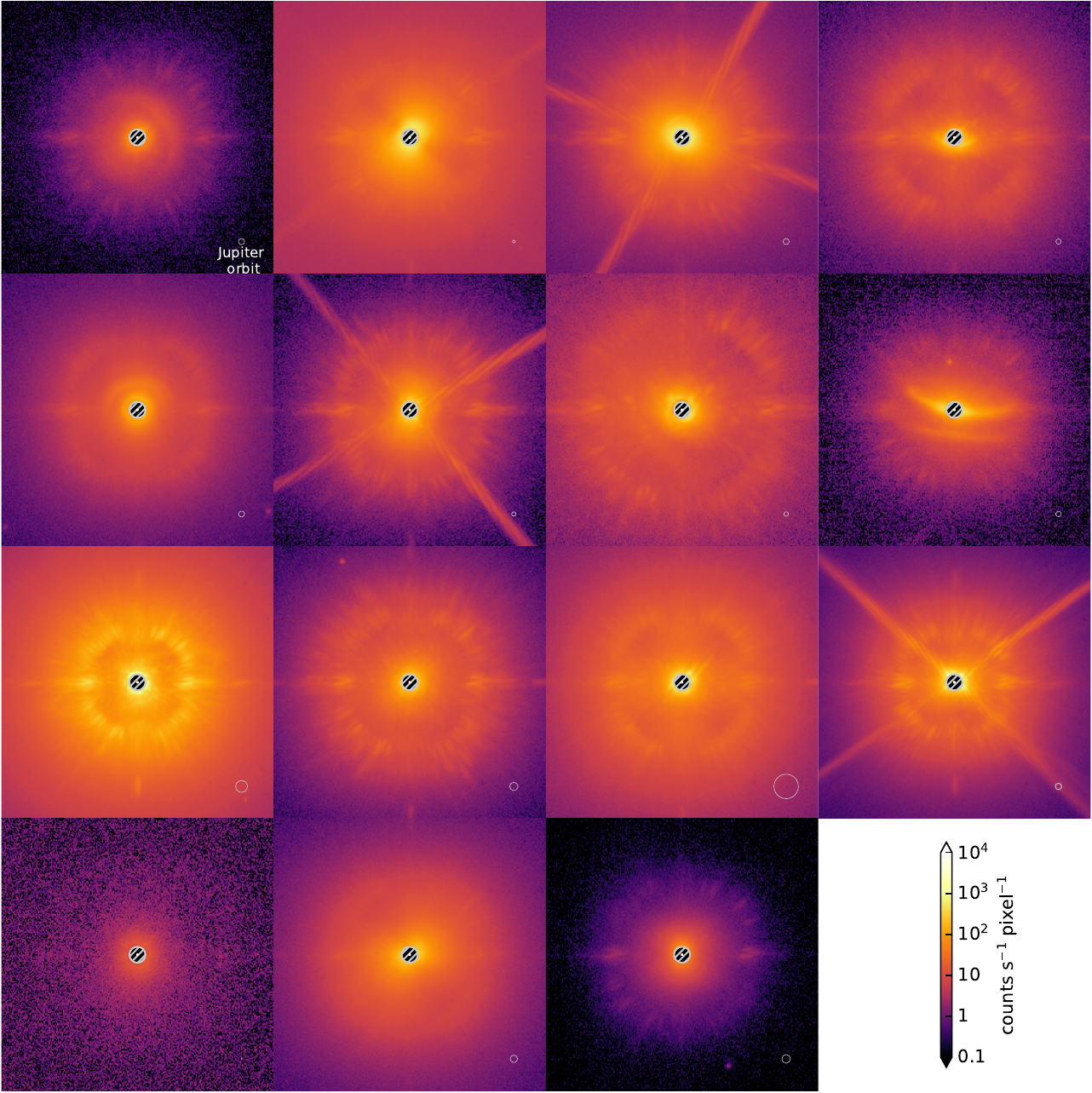}
     \caption{Sample POLARIS preprocessed images, the panels are ones of the the corresponding processed exposures for Figure~\ref{fig-app-gallery}. Some of the brightest disks in Figure~\ref{fig-app-gallery} are marginally seen here. In comparison, the disks here are in total intensity instead of polarized light, see \cite{ren23} for the difference. Notes: (1) Regions interior to the circles (i.e., adapative optics control region) are the regions for data imputation. (2) Some of the exposures have "x"-shaped lines, which are the diffraction spikes of the supporting structure of VLT's secondary mirror.}
     \label{fig-app-gallery-total}    
 \end{figure}

We present manually-selected 15 \Qphi\ files -- from the 921 POLARIS \Qphi\ ones -- to show the morphological diversity of circumstellar structures in Figure~\ref{fig-app-gallery}. For each \Qphi\ image, it is the PDI product of the preprocessed files from \texttt{IRDAP}: we present one of the preprocessed image corresponding to Figures~\ref{fig-app-gallery} and~\ref{fig-app-gallery-total}. The POLARIS files are in \texttt{.fits} format \cite{pence10} following astronomy standards, and thus the figures have no actual color information but just detector counts representing incident light intensity. Most colored plots in this work are based on Python scripts using \texttt{astropy} and \texttt{matplotlib}, where the \texttt{.fits} data are cropped, log-normalized, and visualized with the \texttt{inferno} colormap to enhance morphological features.

Each reference/target system was observed over multiple nights and across different years. The number of total intensity exposures acquired per observation varies depending on the observational strategy, atmospheric conditions, and instrument performance on the specific night of observation. 
In each observation, one can find a \texttt{left.fits} file (ordinary beam) and a \texttt{right.fits} file (extraordinary beam), as products of the dual-beam polarimetry mode of SPHERE (IRDIS DPI Mode). These files are combined to compute the polarized intensity, forming the basis of the POLARIS dataset. In this work, using only the \texttt{left.fits} data for background imputation in Section~\label{A.1} is scientifically justified and validated by the morphology-preserving properties of the dual-beam reduction.

After the background imputation, one should subtract the values from the original image to recover the residuals (i.e. exoplanets or circumstellar disks). The residuals can be rotated using the \texttt{scipy.ndimage.rotate} function's \texttt{angle} input is \texttt{-parangs - 135.99 + 1.75}, where \texttt{parangs} is from the \texttt{*parangs.fits} file for the preprocessed files, to position the images to north-up and east-left, e.g., Figures~\ref{fig-image}b and~\ref{fig-image}c. This geometric alignment ensures that morphological comparisons across systems are performed in a consistent celestial frame.

The primary goals of publishing the POLARIS dataset is to solicit help from the AI community to (1) label the disk-free reference images using \Qphi\ files, and use them to (2) recover the disk- or exoplanet-hosting target images in total intensity. By facilitating automated representation learning and recovery efforts, POLARIS is positioned to accelerate progress in both astrophysical discovery and algorithmic development. In the future, categorizing the reference and target images without the help from \Qphi\ files would further benefit the HCI community.

\begin{table*}[htpb!]
\centering
\caption{Comparing classification accuracy on top of image representations learned from state-of-the-art representation learning methods and Diff-SimCLR.\label{res_2}}
\begin{tabular}{l c c c c c c c c c c}
\toprule 
\textbf{Data} & \textbf{Maskencoder} & \textbf{SimCLR} & \textbf{DeepCluster} & \textbf{Diff-SimCLR}\\ 
\midrule
POLARIS & 85.00 & 86.46 & 74.00 & {\bf 93.00}$^{*}$ \\
\bottomrule
\end{tabular}
\end{table*}

\begin{table*}[htpb!]
\centering
\footnotesize
\setlength\tabcolsep{2pt}
\caption{Performance comparison through classification accuracy among different unsupervised machine learning classifiers across varying representation dimensions. GMM is not performed from dimension of 64 onward due to the risk of overfitting.\label{res_3}}

\begin{tabular}{l | c c c | c c c | c c | c c}
\toprule
& \multicolumn{3}{c|}{\textbf{16-D Features}} & \multicolumn{3}{c|}{\textbf{32-D Features}} & \multicolumn{2}{c|}{\textbf{64-D Features}} & \multicolumn{2}{c}{\textbf{128-D Features}} \\
\textbf{Model} & \textbf{KNN} & \textbf{GMM} & \textbf{Spectral} & \textbf{KNN} & \textbf{GMM} & \textbf{Spectral} & \textbf{KNN} & \textbf{Spectral} & \textbf{KNN} & \textbf{Spectral} \\ 
\midrule
Maskencoder & 
73.78 & \textbf{74.22} & 74.78 & 73.78 & 
74.00 & 77.00 & \textbf{73.78} & 75.89 & 73.78 & 76.78 \\

SimCLR & 
\textbf{75.22} & 72.22 & \textbf{76.33} & 73.89 & 
71.11 & \textbf{77.78} & 70.89 & 71.78 & 67.89 & 70.78 \\

DeepCluster & 
69.89 & 71.00 & 74.78 & 70.67 & 
72.00 & 74.89 & 72.00 & 74.89 & 74.00 & 75.89 \\
\hline
\textbf{Diff-SimCLR} & 
70.56 & 73.22 & 73.56 & \textbf{75.00} & 
\textbf{74.22} & 77.33 & 72.67 & \textbf{76.00} & \textbf{74.78} & \textbf{78.00} \\
\bottomrule
\end{tabular}
\end{table*}
\vspace{3mm}

\begin{table*}[htpb!]
\centering
\footnotesize
\setlength\tabcolsep{2pt}
\caption{Classification accuracy comparison among both supervised and unsupervised machine learning classifiers for Diff-SimCLR representations with varying numbers of latent states embedded from the DDPM.\label{res_4}}

\definecolor{lightpurple}{RGB}{238, 230, 255}
\definecolor{lightgreen}{RGB}{230, 255, 208}

\begin{tabular}{l c c c c c c c}
\toprule 
\textbf{Latent States} & \textbf{SVC} & \textbf{Random Forest} & \textbf{MLPClassifier} & \textbf{SVM} & \textbf{KNN} & \textbf{GMM} & \textbf{Spectral}\\ 
\midrule
$\Delta_t = 2$ & 
\cellcolor{white}88.78 & \cellcolor{white}84.22 & \cellcolor{white}87.50 & \cellcolor{white}86.33 & 
\cellcolor{white}69.78 & \cellcolor{white}76.11 & \cellcolor{white}71.78\\

$\Delta_t = 4$ & 
\cellcolor{white}84.44 & \cellcolor{white}79.00 & \cellcolor{white}84.38 & \cellcolor{white}77.98 & 
\cellcolor{white}68.78 & \cellcolor{white}73.11 & \cellcolor{white}75.00\\
$\Delta_t = 8$ & 
\cellcolor{white}93.00 & \cellcolor{white}89.67 & \cellcolor{white}92.71 & \cellcolor{white}89.56 & 
\cellcolor{white}75.00 & \cellcolor{white}74.22 & \cellcolor{white}77.33\\
$\Delta_t = 16$ & 
80.44 & 79.33 & 81.25 & 84.56 & 
72.78 & 70.00 & 75.00\\
\bottomrule
\end{tabular}
\end{table*}
\vspace{-3mm}

\subsection{Architectural and Hyperparameter Settings}
\label{AppendixA2}

The regularization parameter $\mathcal{C}$ for both SVC and SVM is searched over [0.001, 0.01, 0.1, 1, 10], with SVM kernels selected from [\texttt{rbf}, \texttt{polynomial}, \texttt{linear}]. For RF, tree depth ([5, 10, 15]), minimum samples per leaf, and minimum samples per split are tuned. For MLPClassifier, hidden layer sizes are selected from [10, 20, 30] and learning rates ranged from $1e^{-3}$ to $1e^{-1}$. All search spaces are deliberately constrained to mitigate overfitting. All experiments are conducted on 5 NVIDIA A5000 GPUs using PyTorch.

\begin{figure}[htb!]
\centering
 	\includegraphics[width=\textwidth]{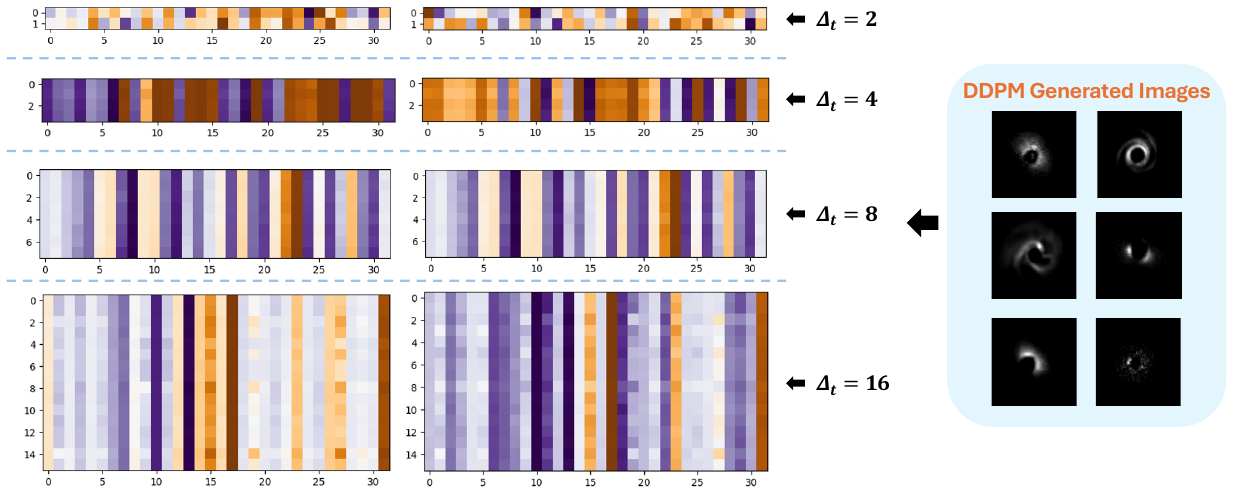}
     \caption{Illustration of the selection of different numbers of the final $\Delta_t$ latent states. The left panel shows a heatmap visualization of the selected latent states for various values of $\Delta_t$. The right panel presents the corresponding generative output from the DDPM at time step $t = T$, representing a high-fidelity, denoised reconstruction of the input image (POLARIS). The generative results for different choices of $\Delta_t$ are shown in a concentrated form, with each sample reflecting the influence of the selected latent subtrajectory.}
     \label{fig-ddpm}    
 \end{figure}

\begin{figure}[htb!]
 	\includegraphics[width=0.97\textwidth]{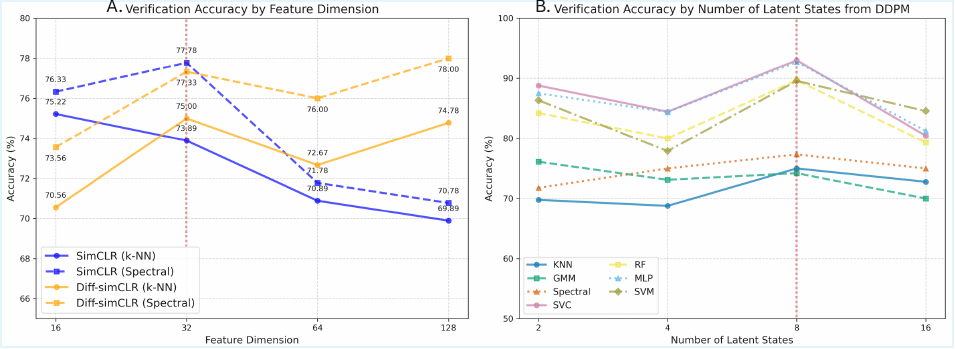}
     \caption{A. Visualization of representation performance learned by contrastive learning-based models corresponding to the results in Table~\ref{res_3}, showing the relationship between feature dimensionality and accuracy on unsupervised downstream tasks. B. Visualization of models from Table~\ref{res_4}, highlighting that the DDPM variant with $\Delta_t = 6$ consistently achieves superior performance across most downstream tasks.}
     \label{fig-line-graph}    
 \end{figure}

The number of latent states from DDPM $\Delta_t$ in Section~\ref{sec:baseline_frams_proposed}, is evaluated over a searching region of [2, 4, 8, 16]. The prior trajectory $p$ has shape $(\Delta_t + 1) \times 32$ and is subsequently encoded using a ResNet-based backbone, as they are the concentrated form of generated reconstruction of POLARIS at states $t = [0, 1, ...., \Delta_t]$ (Figure~\ref{fig-ddpm}). With the Diff-SimCLR output representations fixed as 32-dimensional vectors, we evaluate the impact of different $\Delta_t$ selections by analyzing downstream task performance on the learned representations. For all supervised classifiers (SVC, SVM, Random Forest, MLPClassifier) and unsupervised methods (KNN, GMM, and Spectral Clustering), the respective hyperparameters are consistently optimized using cross-validation-based grid search, as described in Section~\ref{downstream_task_res} and earlier in this section. Table~\ref{res_4} presents the accuracy of the Diff-SimCLR model as a function of the number of latent states. Fewer latent states provide insufficient information to guide contrastive learning, while too many lead the DDPM to capture excessive noise, degrading generative quality and overall performance. This trade-off is illustrated in Figure~\ref{fig-line-graph}, where $\Delta_t = 8$ achieves the best balance. Nonetheless, the variability introduced by retraining and other external factors should be acknowledged.

\begin{figure}[htb!]
\centering
 	\includegraphics[width=\textwidth]{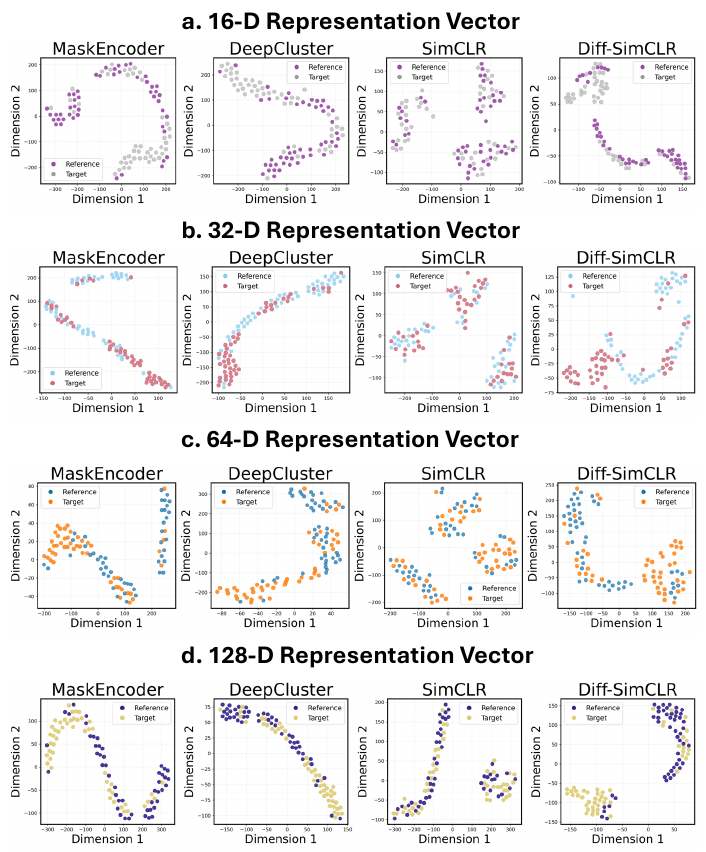}
     \caption{t-SNE visualizations across four models with varying feature dimensions demonstrate that Diff-SimCLR achieves the most distinct and well-separated clusters, indicating stronger representation learning. In contrast, MaskEncoder and DeepCluster produce linear-like feature distributions, while SimCLR shows moderate clustering with less accurate class distinction. }
     \label{fig;representation}
 \end{figure}

Table~\ref{res_2} indicates that our proposed Diff-SimCLR significantly outperforms all baselines on supervised downstream tasks. Moreover, to investigate the impact of representation dimensionality on model performance, we conduct a parallel comparison across feature vector sizes $[16, 32, 64, 128]$ using selected unsupervised downstream tasks. Each model—MAE, DeepCluster, SimCLR, and Diff-SimCLR—is proportionally adjusted to accommodate the respective feature dimensions, either by resizing the autoencoder bottlenecks or scaling the output layers of the ResNet backbones. While higher-dimensional representations generally enhance the expressive power of the learned features, they also increase the risk of overfitting in supervised tasks and may destabilize clustering with GMM, resulting in the exclusion of some configurations at larger dimensions.  As shown in \ref{fig;representation}, the t-SNE visualizations highlight that Diff-SimCLR achieves the most distinct cluster separation across dimensions, indicating more effective representation learning.

Table~\ref{res_3} summarizes the comparative performance. Diff-SimCLR begins to outperform other models from a dimension of 32 onward. At size 16, although SimCLR remains competitive, all models exhibit diminished effectiveness, reflecting a global limitation in expressive capacity at low dimensionality. Notably, Diff-SimCLR underperforms at size 16, likely due to the overhead introduced by latent state embeddings; the added model complexity necessitates aggressive downsampling, which may impair feature quality. A comparative trend is visualized in Figure~\ref{fig-line-graph}A, highlighting a key contrast: while SimCLR degrades with increasing dimensionality, Diff-SimCLR maintains or improves.

\begin{figure}[htb!]
\centering
 	\includegraphics[width=\textwidth]{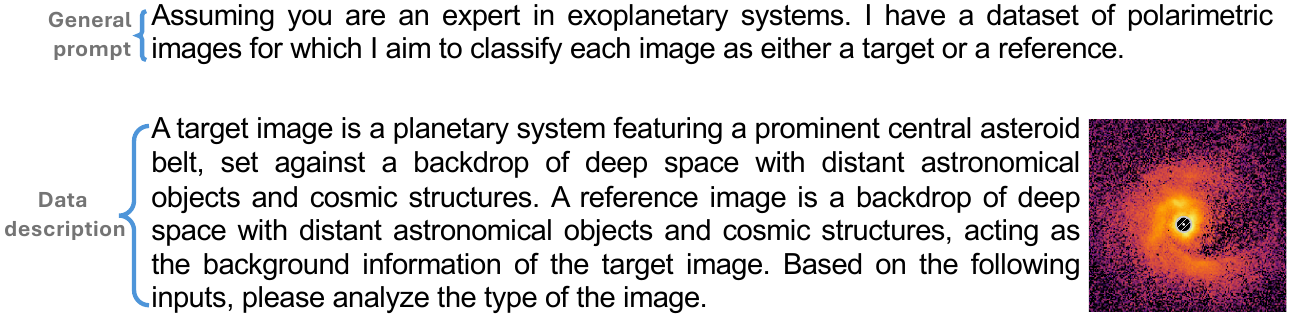}
     \caption{An example prompt for an image of the POLARIS dataset.} 
     \label{llm_prompt}  
     \vspace{-6mm}
\end{figure}

Figure~\ref{llm_prompt} shows an example prompt designed for an image in POLARIS dataset. Our expert-designed prompt consists of two parts: (i) the general prompt which introduces the task scenario and (ii) dataset description which describes the characteristics of the target and reference images we want to focus on.

\subsection{Extended Results and Future Directions for Disk Reconstruction}

The VAE model proposed in Section~\ref{downstream_task_res} contains a convolutional autoencoder containing 3 layers converts raw images to a $128 \times 32 \times 32$ tensor, then fully connects to 4 encoding keys. 2 keys represent the Gaussian distribution parameters, and the other 2 keys represent the scaling factor and bias. We adapt the loss functions to learn spatial patterns in images. The MSE loss is used for light intensity regression, ensuring that the overall brightness of the imputed image aligns with the target background image. KL regularization loss is applied to mitigate image-to-image variance. Additionally, the sum of four MSE losses computed on post-convolutional features is used to enhance pattern learning. The total training loss is the weighted sum of all the losses described above. Extended reconstruction results for selected systems are presented in Figure~\ref{fig-cluster-vae}, demonstrating the effectiveness of classification performed on learned representations in guiding VAE-based reconstructions.

\begin{figure}[htb!]
 	\includegraphics[width=\textwidth]{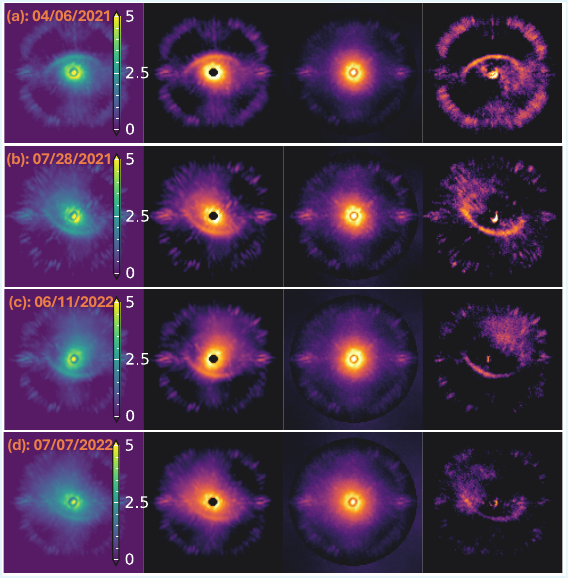}
     \caption{VAE results across multiple epochs of HD~163286, shown in time order from (a) to (d), with layout consistent with Figure~\ref{fig-cluster-vae}.}
     \label{fig:vae-new}    
 \end{figure}

Figure~\ref{fig:vae-new} presents multiple exposures of the same system, HD 163286, captured at different epochs, illustrating the temporal variability inherent in such observations. While the current method processes single-frame ADI data meaningfully, future work should address temporally coherent, multi-epoch sequences that exhibit periodic motion due to pupil tracking. Such structured data necessitate generative models with embedded physical constraints, as standard generative models (i.e. VAE) would fail to preserve the underlying astrophysical dynamics. To ensure physical plausibility, future work should incorporate continuous spatial encoding and domain-specific constraints within generative models. Such models have the potential to enhance both data-driven AI modeling in astrophysics and the broader development of astronomical grounded generative learning frameworks.

\vspace{-6mm}

\newpage

 \newpage

   \begin{figure}[htb!]
\centering
 	\includegraphics[width=0.85\textwidth]{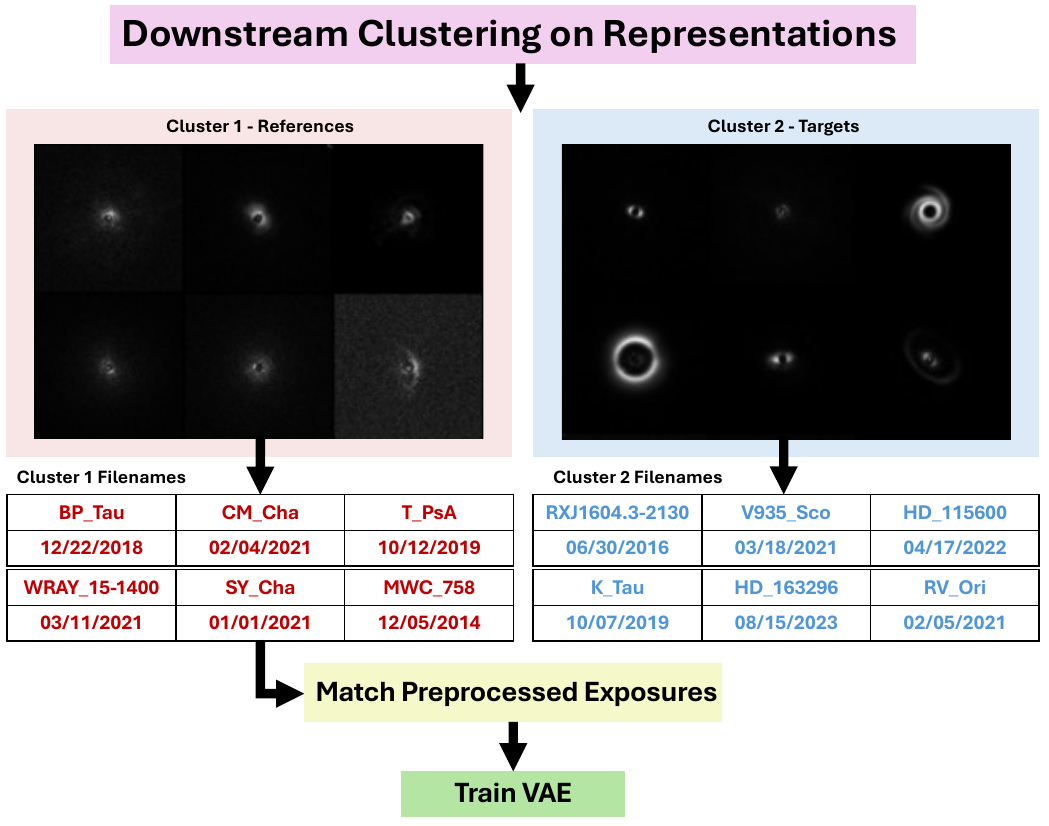}
     \caption{ The representations of polarized images learned using Diff-SimCLR are utilized in downstream classification tasks—specifically, spectral clustering—to identify two distinct clusters (corresponding to known labels or reference categories), each associated with a particular system and its observation time. The resulting clustering is also shown in Figure~\ref{fig:unsupervised_clustering}. Based on this clustering, the corresponding preprocessed exposures, specifically the RDIs, are selected for training the VAE model. This model is then used to learn the distribution of the stellar PSF, facilitating the reconstruction of circumstellar disk structures.}
     \label{fig-cluster-vae}    
 \end{figure}

\newpage

   \begin{figure}[htb!]
\centering
 	\includegraphics[width=\textwidth]{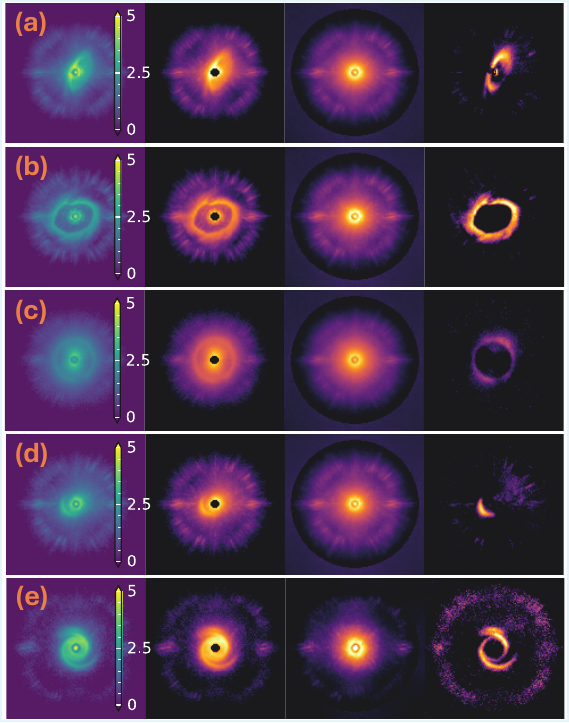}
     \caption{VAE results from the selected model described in Section~\ref{downstream_task_res}. Each column (left to right) shows: (1) the raw exposures, (2) the preprocessed exposures represented in total light intensity, (3) the VAE-predicted stellar background (i.e., starlight component), and (4) the residual image highlighting the exoplanetary or circumstellar disk emission after background subtraction. Each row (top to bottom) corresponds to a different target:  
        (a) V351 Orionis (HD~38238), identical to the observation shown in Figure~\ref{vae_disk};  
        (b) HD~37400;  
        (c) J1604 (2MASS~J16042165--2130284);  
        (d) V1247 Orionis (HD~290764);  
        (e) HD~36112.\\
        \textit{Note}: The diffuse outer ring of emission and the ``$\times$''/``$+$'' shaped patterns are artifacts caused by instrumental mirror-related issues as mentioned in Figure~\ref{fig-app-gallery-total}.
    }

     \label{fig-cluster-vae}    
 \end{figure}

\end{document}